\documentclass[11pt,a4paper]{article}
\usepackage{jheppub}

\usepackage{latexsym}
\usepackage{graphics}
\usepackage{pstricks}
\usepackage{epstopdf}
\usepackage[T1]{fontenc}
\usepackage{amsmath}
\usepackage{amsfonts}
\usepackage{amssymb}
\usepackage{empheq}
\usepackage[utf8]{inputenc}
\usepackage{graphicx}
\usepackage{epsfig}
\usepackage{lineno}
\usepackage{color}

\usepackage[poly,arrow,curve,matrix]{xy}
\usepackage{pst-node}
\usepackage{tikz-cd} 
\usepackage{tikz}
\usetikzlibrary{shapes.geometric, arrows}
\usepackage{mathptmx}
\tikzset{>=latex}

\newtheorem{theorem}{Theorem}[section]

\title{{\bf
Boundary-to-bulk maps for AdS causal wedges and  RG flow}}

\author[a,b]{Nicolás Del Grosso} 
\author[a,b]{Alan Garbarz} 
\author[a]{Gabriel Palau}
\author[a]{Guillem Pérez-Nadal}

\affiliation[a]{Physics Department, University of Buenos Aires\\ 
Pabell\'on 1, Ciudad Universitaria, 1428 Buenos Aires, Argentina}
\affiliation[b]{IFIBA-Conicet\\ 
Pabell\'on 1, Ciudad Universitaria, 1428 Buenos Aires, Argentina}

\emailAdd{ngrosso@df.uba.ar}
\emailAdd{alan@df.uba.ar}
\emailAdd{gbrlpl8@gmail.com}
\emailAdd{guillem@df.uba.ar}

\abstract{We consider the problem of defining spacelike-supported boundary-to-bulk propagators in AdS$_{d+1}$ down to the unitary bound $\Delta=(d-2)/2$. That is to say, we construct the `smearing functions' $K$ of HKLL but with different boundary conditions where both dimensions $\Delta_+$ and $\Delta_-$ are taken into account. More precisely, we impose Robin boundary conditions, which interpolate between Dirichlet and Neumann boundary conditions and we give explicit expressions for the distributional kernel $K$ with spacelike support. This flow between boundary conditions is known to be captured in the boundary by adding a double-trace deformation to the CFT. Indeed, we explicitly show that using $K$ there is a consistent and explicit map from a Wightman function of the boundary QFT to a Wightman function of the bulk theory. In order to accomplish this we have to study first the microlocal properties of the boundary two-point function of the perturbed CFT and prove its wavefront set satisfies the microlocal spectrum condition. This permits to assert that $K$ and the boundary two-point function can be multiplied as distributions. 
}

\begin{document}



\maketitle



\section{Introduction}

The framework of AdS/CFT in Lorentzian signature made significant progress when Hamilton et al. showed how to reconstruct bulk local operators from CFT primary operators, at least in the large N limit \cite{Hamilton:2005ju,Hamilton:2006az}. This is important because it allows to probe locality of correlators in the bulk in terms of correlators of the boundary CFT evaluated on points that are causally connected. In those references a `smearing function' $K$ was found such that it has spacelike support and allows to write a bulk field $\Phi$ in terms of its boundary data $\phi$ smeared over the boundary:
\begin{equation}\label{Kmap}
\Phi(X)=\int_{\Omega(X)} d^dx \sqrt{-h} K(X,x) \phi(x) 
\end{equation}
Here $h$ is the boundary metric, $\Omega(X)$ is the region in the boundary that is spacelike separated from $X$, and 
$$\phi(x)=\lim_{z\rightarrow 0} z^{-\Delta}\Phi(z,x)$$
for some suitable radial coordinate $z$ that approaches the boundary as tends to zero. The dimension of $\phi$ is $\Delta_+\geq d/2$. In this work we are interested in reformulating this Lorentzian bulk reconstruction to allow the dimension to reach the unitarity bound $(d-2)/2$, which amounts to consider mixed boundary conditions in the bulk, and  a relevant deformation of the CFT. We are going then to discuss the necessary generalization of $K$ and explicitly show that it relates correlators on the boundary QFT with correlators on the bulk. The more interesting scenario with interactions was considered later \cite{HKLLinteractions1,HKLLinteractions2}, however in this work we will stick to the free linear theory.

It is a simple fact, already observed in \cite{Hamilton:2005ju}, that $K$ is not unique, since one can always add to $K$ a kernel $K'$ orthogonal to $\phi$ in the sense\footnote{As explained on footnote 7 of \cite{Morrison}, the reason behind the non-uniqueness is that if $K$ is regarded as acting on CFT correlation functions, then as these correlations do not have spacelike support on momentum space, $K'$ can be anything with Fourier trasform with spacelike momenta.} $\int_\Omega K' \phi=0$. Actually the authors used this freedom to construct a real $K$. The existence of $K$, on the other hand, is a rather subtle issue. Although it is sometimes mentioned that $K$ does not exist in certain situations, for example when localized on the AdS-Rindler patch \cite{Hamilton:2005ju, Hamilton:2006az, Leichenauer:2013kaa}, one needs to specify what sort of mathematical object $K$ is considered to be. As a function, it certainly does not exist. As an integral kernel used to map boundary operators to bulk operators, it is claimed not to exist in general in \cite{Bousso:2012mh,Leichenauer:2013kaa}. In those references, the non-existence conclusion follows from first considering the bulk field and boundary operator both  decomposed in modes
$$\Phi(X)=\int d^dk\, a_k F_k(X), \quad \phi(x)=\int d^dk\, a_k f_k(x)$$
where we are using that $F_k(z,x)\sim z^{-\Delta} f_k(x) $ for small enough $z$ (i.e., close to the boundary). Then, assuming\footnote{In \cite{Leichenauer:2013kaa} every case studied satisfies this assumption.} that both $F_k$ and $f_k$ form orthonormal basis, one can write the amplitudes as $a_k=\int d^dx\sqrt{-h}\, f^*_k \phi $, the bulk field reads $\Phi(X)=\int dk \, \int d^dx \sqrt{-h} f^*_k \phi F_k(X) $. Then further assuming the order of integration can be exchanged, the kernel reads
\begin{equation}
K(x|X)=\int dk\,  f^*_k(x) F_k(X)
\end{equation}
The modes $F_k$ will typically diverge for large enough $k$ and such integral does not converge. The exceptions are global AdS and the Poincaré patch (which has polynomial growth).    

However, the analysis of Morrison in \cite{Morrison} shows that one has to be more careful and define what $K$ is supposed to do. One goal is to use $K$ as a map from correlation functions of the CFT to correlation functions in the bulk theory, and a more ambitious goal is to use $K$ to map a local bulk operator algebra to a local boundary operator algebra. Then, for example one important issue is whether the two-point function  $\langle \Phi(X_1) \Phi(X_2)\rangle$ can be written as (from now on we omit the boundary metric determinant) 
$$\langle \Phi(X_1) \Phi(X_2)\rangle=\int_{\Omega(X_1)\ \cup\ \Omega( X_2)} d^dx_1\,d^dx_2\, K(x_1|X_1)K(x_2|X_2) \langle \phi(x_1) \phi(x_2) \rangle $$ 
To answer such a question one has to notice that distributions are being multiplied. One powerful tool that can be used to understand if this multplication of distributions makes sense is the wavefront set of the distributions (see for example \cite{Hormander, Brouder:2014hta}). Roughly speaking, the multiplication between $K$ and $\langle \phi \phi \rangle $ can be defined if the problematic directions in momentum space of these distributions are not equal and of opposite sign (so if $k$ is a `bad' direction of $K$, then it should be that $-k$ is not bad for the boundary correlator). Morrison showed that even in the case of a causal wedge in AdS, where the modes grow exponentially, this dangerous behavior only occurs in directions where the correlator can be regarded as a smooth function\footnote{The technical assumption in \cite{Morrison} is that the two-point function of the boundary CFT satisfies the microlocal spcetrum condition of \cite{Brunetti:1995rf}. We will review this assumption in Section \ref{Mapping}.}, and then the multiplication of both distributions is well defined. 

We can be a little bit more explicit on how we see $K$, following \cite{Morrison}. The bulk field $\Phi$ is a distribution that takes compactly supported smooth test functions $F$ and gives back operators $\Phi[F]$ on some algebra (typically a unital $*-$algebra). The Klein-Gordon equation of motion in the bulk is satisfied in the sense that $\Phi[(\square -m^2)F]=0$ for any $F$. On the boundary CFT we consider an algebra of operators $\phi[f]$ labeled by test functions $f$. However, in order to get the \textit{local} algebra of the bulk from the CFT algebra, the test functions of the CFT need to be quite general. More precisely, as mentioned above,  we want to be able to write,
\begin{equation}\label{twopointfunctionholography}
\langle \Phi[F] \Phi[F']\rangle =\int_{\Omega(\text{supp}(F))\ \cup\  \Omega(\text{supp}(F'))}d^dx\,d^dx' K[x|F]K[x'|F'] \langle \phi(x) \phi[x']\rangle,\quad \forall \,F \,\text{and} \,F'
\end{equation}  
where $K[F,\cdot]=\int_{\text{supp}(F)} K(\cdot|X) F(X)$ is a function on the boundary and in virtue of the above expression should be taken to be a boundary test function: 
\begin{equation}\label{bdytestfunctions}
f_F:=\int_{\text{supp}(F)}d^{d+1}X\ \sqrt{-g} K(\cdot|X) F(X)=K[\cdot|F]
\end{equation}
Notice, however, that this object could fail to be an actual function, since as discussed earlier $K$ could have expontial growth in Fourier space and this expression would be ill-defined. But the point made in \cite{Morrison} is that $f_F$ makes sense inside an integral, such as in (\ref{twopointfunctionholography}). Then, the space of boundary test functions has to be enlarged in order to accomodate those $f_F$ that are actually distributions of the form (\ref{bdytestfunctions}). moreover, the product of ditributions  in (\ref{twopointfunctionholography}) of the form $f_F(x) \cdot \langle \phi(x) \phi(x')\rangle$ needs to be well defined, and in \cite{Morrison} a sufficient condition for the wavefront set WF$(f_F)$ is given. In this sense, even in AdS causal wedges, $K$ exists despite its exponential growth in momenta and the `test functions' $f_F$ turn out to be\footnote{In short, since the two-point function has singularities when $x$ and $x'$ are null separated and in directions of locally-positive frequency, the sufficient condition for $f_F$ is that its singular directions in momentum space are spacelike. In this way, $f_F$ behaves as a smooth function in null and timelike momentum directions and can be multiplied with the two-point function.\label{footnote:prod}} sufficiently well behaved as distributions. 
 
As already mentioned, we are interested in extending this analysis beyond the Dirichlet boundary condition. Put differently, we would like to test the HKLL holographic construction \textit{even when the boundary theory is not conformal.} To this end, we start by considering the fact that AdS is not globally hyperbolic, which implies that boundary conditions at timelike infinity need to be imposed in order to have a well-defined evolution of initial conditions. Different boundary conditions have been explored in the literature \cite{Avis:1977yn,Ishibashi:2004wx} and are well understood from a bulk point of view (see \cite{Dappiaggi:2018xvw} for a thorough treatment). One main point is that the behavior of the scalar field can have both $z^{\Delta_+}$ and $z^{\Delta_-}$ decays in the Breitenlohner-Freedman (BF) window $0\leq \nu \leq 1$ \cite{Breitenlohner:1982bm}, where $\Delta_\pm = d/2 \pm \nu $, and $\nu=\sqrt{d^2/4+m^2}$. Then, our first step is to develope a holographic reconstruction of the bulk dynamics with mixed boundary conditions. In other words, we want to reformulate the work initiated by Hamilton et al. in \cite{Hamilton:2005ju,Hamilton:2006az} in order to account for the freedom in choosing different boundary conditions from the usual Dirichlet boundary condition ($\Phi\sim z^{\Delta_+}$) . In particular, based on the earlier results of \cite{Ishibashi:2004wx} and the recent work of \cite{Dappiaggi:2018xvw}, we consider Robin boundary conditions. Immediately we will see the need to proceed with care, since divergences appear when imposing Neumann boundary conditions ($\Phi\sim z^{\Delta_-}$), at least when working in position space. In order to be more precise, let $d=1$  so the bulk-to-boundary smearing function $K$ of HKLL, when evaluated at the origin of AdS$_2$, behaves as 
$$K \sim \cos(t)^{\Delta_+-1} \Theta(t-\pi/2)\Theta(t+\pi/2)$$
where $\sim$ means we are ignoring some normalization constants for the moment. Now, the BF window  is $-1/4\leq m^2 \leq 3/4$ and within this range we can instead take Neumann boundary conditions. All we have to do at this point is to replace $\Delta_+$ with $\Delta_-$. This gives $K\sim\cos(t)^{-\Delta_+}$. Since $1/2 < \Delta_+ <3/2$, 
$$ \Phi(0) \sim \int_{-\frac{\pi}{2}}^{\frac{\pi}{2}} dt \cos(t)^{-\Delta_+} \phi(t) $$ 
is not in general convergent for $1<\Delta_+<3/2$. 

This kind of behavior already appears in the much studied Dirichlet case for $d>1$. Take again the origin of $AdS_{d+1}$ with odd $d$, and we have $K \sim \cos(t)^{\Delta_+-d}$. Then one needs to require $\Delta_+ > d-1$. However this bound has no physical significance and for any $d>1$ this is too stringent, so it would be desirable to have a smearing $K$ for the range $d/2<\Delta_+<d-1$. This is easy to accomplish by means of an analytic continuation similar to the one used when defining the $\Gamma$ function in the entire complex plane \footnote{The procedure follows the same lines as the extension of the generalized  function $x_+^{\lambda}$  for $\lambda<-1$ and $\lambda\neq -n$ \cite{Gelfand}. We shall come back to this observation later.}. 

We are now in a position to state our strategy in order to construct $K$ for Robin boundary conditions: looking at a bulk field with Dirichlet boundary conditions and weight $\Delta_+$ as a function of $\Delta_+$, $\Phi_{D}(\Delta_+)$, we can relate it to its analogue for Neumann boundary conditions by just exchanging $\Delta_+$ with $\Delta_-$
$$\Phi_N={\Phi_D}|_{\Delta_+\mapsto \Delta_-}$$
This simple step, as commented above, calls for care when viewing K as an integral kernel. This is because we are performing the replacement $\nu \rightarrow -\nu$ in $K$ and the integrand of $\int K \phi$ becomes more divergent near the limits of integration. However, by analytically continuing $K$ in $\nu$ we will be able to define $K$ for Neumann boundary conditions. Then, the smearing function $K$ for Robin boundary conditions is just an appropriate linear combination of the Dirichlet and Neumann $K$'s. The  distribution $K$ will not be an integral kernel, though, since it will have delta-like contributions.        

So far we have discussed the continuation of the HKLL smearing function to Neumann boundary conditions in global AdS and in its coordinate representation. This representation allows comparision with the original results of \cite{Hamilton:2005ju,Hamilton:2006az}. We should say, however, that when considering AdS causal wedges (in particular the Poincaré patch), there is a very natural representation of $K$ as a Fourier transform of the boundary coordinates and which allows to make the analytic continuation in a straightforward way. In other words, the replacement $\nu\rightarrow -\nu$ requires no care if it is performed on the momentum space representation of $K$. We will justify this claim and take advantage of the Fourier-transform presentation of $K$ when we consider the Poincaré and Rindler patches. In particular it will allow us to show that $K$ is spacelike supported. 

Next we would like to discuss the counterpart on the boundary side, which comes from considering multi-trace deformations of the CFT \cite{KlebanovWitten,Wittenmulti}. In particular, relevant double-trace deformations by operators with dimension $\Delta<d/2$ of the CFT generate an RG-flow from a UV fixed point (Neumann) to an IR fixed point (Dirichlet) (see also  \cite{GubserKlebanov, GubserMitra, HartmanRastelli}). These observations were made in the context of the standard AdS/CFT dictionary \cite{Maldacena:1997re}, namely relating the Euclidean path integrals of the bulk and boundary theories \cite{Gubser:1998bc,Witten:1998qj}. Here we will show that the HKLL point of view can still be applied when an RG flow takes place at the boundary and conformal invariance is broken. In particular we will show two things. First, that the (very reasonable) assumption of \cite{Morrison} about the microlocal spectrum condition of the boundary two point function is correct for any point of the RG flow (in particular the IR CFT considered in that reference). This allows to claim that the product as distributions of $K$ and the boundary correlator is well defined (as explained in footnote \ref{footnote:prod}). Second, we will explicitly map the  boundary two-point function (of the deformed CFT) to the bulk two-point function with mixed boundary conditions, and obtain agreement with the one previously obtained in \cite{Dappiaggi:2016fwc}.     

The manuscript is organized as follows. In Section \ref{KGReview} we review the analysis of solutions of the Klein-Gordon equation on AdS$_{d+1}$ with Robin boundary conditions mainly from a (singular) Sturm-Liouville theory point of view, but we make an effort to connect with the usual  presentations. In Section \ref{HKLLglobal} we make first a revision of the original results of HKLL in global AdS and then extend them to Robin boundary conditions. In Section \ref{HKLLwedges} we revisit the results of HKLL and Morrison and extend them to Robin boundary conditions as well as proving the spacelike support property. The reader mainly interested in the mapping between boundary and bulk correlators can jump straight to Section \ref{Mapping} where this is discussed. We end with a Conclusions section. There is also included an Appendix on wavefront sets of distributions and oscillatory integrals, where we make an extremely brief (but hopefully useful) exposition of these topics and where we prove that the boundary Wightman 2-point function of the perturbed CFT satisfies the microlocal spectrum condition.  

\section{Short review of Klein-Gordon modes in AdS}\label{KGReview}

The bulk solutions to the Klein-Gordon equation come in pairs, as this is a second order differential equation. After a decomposition in spherical harmonics of the $S^{d-1}$ sphere and a Fourier transform in time\footnote{In this section we will concentrate in the radial part of the bulk field $\Phi$. However in the remaining sections we will be interested in the dependence of the field on all the bulk coordinates and we will still call it $\Phi$, in order not to introduce many different notations.}, a singular Sturm-Liouville problem appears for  the radial part \cite{Dappiaggi:2018xvw}. The radial solutions regular at the origin $\rho=0$ are\footnote{Here and in the rest of this work we use the same coordinates as in \cite{Hamilton:2006az}. In short, the AdS radius is set to $1$ and $\rho$ is a radial coordinate where $\rho=0$ is the origin of AdS and $\rho=\pi/2$ is the conformal boundary. The conformal factor of the metric is $(\cos\rho)^{-2}$.}
\begin{eqnarray}\label{Phi1Phi2}
\Phi_1(\rho)&=& (\sin\rho)^{l}(\cos\rho)^{\Delta_+} {}_{2}F_1(a,b;c;\sin^2\rho)\nonumber\\
\Phi_2(\rho)&=& (\sin\rho)^{2-d-l}(\cos\rho)^{\Delta_+} {}_{2}F_1(a-c,b-c;2-c;\sin^2\rho)
\end{eqnarray}
while the regular solutions at the boundary $\rho=\pi/2$ are
\begin{eqnarray}\label{Phi3Phi4}
\Phi_3(\rho)&=& (\sin\rho)^{l}(\cos\rho)^{\Delta_+} {}_{2}F_1(a,b;a+b+1-c;\cos^2\rho)\nonumber\\
\Phi_4(\rho)&=& (\sin\rho)^{l}(\cos\rho)^{\Delta_-} {}_{2}F_1(a-c,b-c;c-a-b+1;\cos^2\rho)
\end{eqnarray}
In these expressions we have used $l$ which is a natural number related to the spherical harmonics and will not be important, and we have omitted the dependence of the $\Phi$'s on $\omega$ (coming from the Fourier transform in global time). Also
\begin{eqnarray}
a&=&\frac{l+\Delta_+-\omega}{2}\nonumber\\
b&=&\frac{l+\Delta_++\omega}{2}\nonumber\\
c&=&l+\frac{d}{2}
\end{eqnarray}
These pairs are lineary independent as long as the third argument in the hypergeometric function is not a natural number. For instance, if $c\in\mathbb{N}$ then another $\Phi_2$ solution appears, but we will not take this into account since $\Phi_2$ is never square-integrable in the Sturm-Liouville problem at hand. The situation for the pair $(\Phi_3,\Phi_4)$ is the most important one for our study. As is well-known, $\Phi_4$ is not square-integrable for $\nu\geq1$, while both solutions are admissible (i.e. they are square-integrable) in the range $0<\nu<1$. This is the case we are interested in, where both weights, $\Delta_+$ and $\Delta_-$, appear.

The way to implement boundary conditions at $\rho=\pi/2$, according to the singular Sturm-Liouville theory \cite{Zettl}, is to demand that the sought solution $\Phi_\gamma$ satisfies for given $\gamma\in[0,\pi)$  
\begin{equation}\label{RobinBC}
\lim_{\rho\rightarrow\frac{\pi}{2}}\ (\tan\rho)^{d-1} \left(\cos\gamma\ W[\Phi_\gamma,\Phi_3] + \sin\gamma\ W[\Phi_\gamma,\Phi_4]\right)=0
\end{equation}
where $W[f,g]=f g'-g f'$ is the Wronskian. We see that we have a one-parameter family of boundary conditions, called Robin boundary conditions. This expression has the virtue of putting the decays $(\cos\rho)^{\Delta_+}$ and $(\cos\rho)^{\Delta_-}$ on equal footing. Let us motivate the boundary condition (\ref{RobinBC}) from a more physical point of view: generically the solutions near the boundary are of the form 
\[ \Phi \sim  (\cos\rho)^{\Delta_+}\phi_+ + (\cos\rho)^{\Delta_-}\phi_- \]
In the simple massless case in $1+1$ dimensions we see that $\phi_-$ is the boundary value of $\Phi$ while $\phi_+$ is the normal derivative of $\Phi$ at the boundary (since $\Delta_+=1$ and $\Delta_-=0$). In general this is not precisely the case, but is still true that since $\Delta_+>\Delta_-$ then $\phi_-$ is the leading value of $\Phi$ close to the boundary. Even more, with the definition $\phi=\lim_{\rho\rightarrow\pi/2}(\cos\rho)^{-\Delta_-} \Phi $, then $\phi=\phi_-$. This is why $\phi_-=0$ is usually called the Dirichlet boundary condition, and $\phi_+=0$ the Neumann boundary condition. Now, it is then natural to write a mixed boundary condition as
\[ \phi_- + c \phi_+ =0 \]    
but this is exactly what one gets from (\ref{RobinBC}), with $c=-\tan\gamma$. In other words we have
\[\phi_+=\cos\gamma\ \phi,\quad \phi_-=\sin\gamma\ \phi \]
with $\phi$ a constant, or if we take into account the other coordinates of the bulk field, $\phi$ is the field at the boundary. 
The advantage of (\ref{RobinBC}) is that it is written in terms of the solutions of the differential equation, and not of asymptotic values.  

The solution to (\ref{RobinBC}) is the linear combination,
\begin{equation}
\Phi_\gamma(\rho)=\cos\gamma\ \Phi_3(\rho) +\sin\gamma\ \Phi_4(\rho)
\end{equation} 
In particular, Dirichlet boundary condition means $\gamma=0$ and the solution is $\Phi_3$, while $\gamma=\pi/2$ should be called the Neumann boundary condition and the corresponding solution is $\Phi_4$.  

Regularity at the origin implies that $\Phi_\gamma$ is proportional to $\Phi_1$, but at the same time $\Phi_1$ can be written as a specific linear combination of $\Phi_3$ and $\Phi_4$ (just because they span the space of solutions), and so a discrete set of frequencies $\omega$ are allowed for given $\gamma$. Say that $\Phi_1=A_\omega \Phi_4 + B_\omega \Phi_3$, then the condition on the set of allowed frequencies is
\begin{equation}\label{Robinfreq}
\tan\gamma = \frac{A_\omega}{B_\omega}
\end{equation}

In \cite{Hamilton:2005ju,Hamilton:2006az} it was used that the frequencies corresponding to Dirichlet come equispaced by $2n$, with $n$ an integer number \cite{Balasubramanian:1998sn}, and this implied immediately the spacelike support of $K$. Such a simplification does not occur with Robin boundary conditions, since the frequencies are given by the transcendental equation above  and are not equispaced (see \cite{Ishibashi:2004wx} as well as \cite{Dappiaggi:2018xvw}). However by the analytic continuation method we will employ, it will be easy to construct $K$ for Neumann boundary condition $\gamma=\pi/2$ and then for generic Robin boundary conditions, and we will see they have spacelike support.   

\section{HKLL map on global AdS reloaded}\label{HKLLglobal}

In this section we revisit the results of \cite{Hamilton:2005ju,Hamilton:2006az} and generalize them to include Robin boundary conditions. We will try here to clearly indicate where their work may need further analysis, such as in the case $d/2<\Delta_+<d-1$ (which includes the Dirichlet condition).  We  skip the derivations and turn to the expressions of the kernel $K$. From now on $\Phi$ denotes the complete bulk field, not just its radial part.

\subsection{HKLL revisited}

In this subsection we are considering kernels that were found in \cite{Hamilton:2005ju,Hamilton:2006az} using a method sketched in the introduction. They act on the boundary value $\phi$ of the Dirichlet bulk field $\Phi_D$. Such boundary value is defined as,
\begin{equation}
\text{Global coordinates:} \qquad  \phi:=\lim_{\rho\rightarrow \pi/2} \frac{\Phi_D}{(\cos\rho)^{\Delta_+}}
\end{equation}

\subsubsection*{$1+1$ dimensions}

Let us begin with AdS$_2$ and the kernel with support on the right boundary, 
\begin{equation}
\label{eq:Kglobal1+1}
\begin{aligned}
K_D(t| t', \rho'):&= \frac{\Gamma(\Delta_+ +\frac{1}{2})\ 2^{\Delta_+ -1}}{\sqrt{\pi}\ \Gamma(\Delta_+)} \lim_{\rho \to \pi/2}\left(   \sigma \ \cos \rho \right)^{\Delta_+ -1} \theta(\rho-\rho'-|t-t'|)\\
&= \frac{\Gamma(\Delta_+ +\frac{1}{2})\ 2^{\Delta_+ -1}}{\sqrt{\pi}\ \Gamma(\Delta_+)} \left( \frac{\cos(t-t')-\sin \rho'}{\cos \rho'}  \right)^{\Delta_+ -1} \theta(\pi/2-\rho'-|t-t'|)
\end{aligned}
\end{equation}
where we have used that 
\begin{equation}
\sigma=\frac{\cos(t-t')-\sin \rho' \sin\rho}{\cos\rho \cos \rho'} 
\end{equation}
This kernel behaves like a compactly supported boundary  distribution around $t'$, and  for $\rho'\rightarrow\pi/2$, i.e. when the bulk point approaches the boundary, the support shrinks. Notice also that the integral $\int dt K \phi(t)$ converges for bounded $\phi$ since, for $d=1$,  $\Delta_+ -1$ is always greater than $-1$ (it is actually greater or equal than $-1/2$). As mentioned in the Introduction, we can see that with the replacement $\nu\rightarrow-\nu$ we will have instead  an integrand of the form $\sigma^{\Delta_- - 1}=\sigma^{-1/2-\nu}$, which is not guaranteed to converge for any bounded $\phi$ unless $\nu<1/2$. However the unitary bound is $\nu=1$. We will see how to make this work shortly. 

An interesting particular case is the massless scalar field, where we have,
\begin{equation}\label{eq:sinmasa}
\Phi_{D}^{M=0}( t', \rho')=\frac{1}{2} \int \limits_{-\pi/2+\rho'+t'}^{\pi/2-\rho'+t'}\ \phi(t)\ dt\ ,
\end{equation}
Thus we see that every boundary point, spacelike separated from the bulk point, contributes equally. We will see that in the case of Neumann boundary conditions this behaviour is drastically changed.
  
\subsubsection*{Even bulk dimensions}

\begin{figure}[h!] 
\centering
    \includegraphics[width=.5\linewidth]{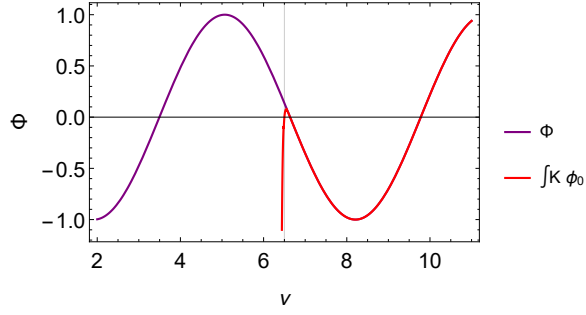} \\
 \caption{We show one mode of $\Phi$ as a function of $\nu$ and the expression $\int K \phi$, for $d=15$. It is evident they coincide only for $\nu>13/2$.  } {\label{fig:badK}}
\end{figure}

In general even bulk dimensions the boundary is connected and the kernel reads,
\begin{equation}\label{eq:Kglobaleven}
    K_D(x|X')=C(\Delta_+) \lim_{\rho\rightarrow\pi/2}(\sigma(x|X') \cos \rho)^{\Delta_+-d}\ \Theta(\sigma-1) \ .
\end{equation}
where, 
\begin{equation}
\sigma=\frac{\cos(t-t')-\sin \rho' \sin\rho \cos\left(\Omega-\Omega'\right)}{\cos\rho \cos \rho'} 
\end{equation}
and
\begin{equation}\label{eq:Ceven}
C(\Delta_+)=\frac{(-1)^{(d-1)/2}2^{\Delta_+-d-1}\Gamma(\Delta_+-d/2+1)}{\pi^{d/2}\Gamma(\Delta_+-d+1)}
\end{equation}
Here $\Omega-\Omega'$ represents the angular separation in the $d-1$-dimensional sphere. In this case, as opposed to $d=1$, the integral kernel is only defined for $\Delta_+>d-1$, so there is a window $d/2<\Delta_+<d-1$ where the original HKLL does not make sense. In Figure \ref{fig:badK} this is shown for $d=15$, where the kernel breaks down at $\nu=13/2$ and below.  Again, we will solve this issue at the same time of changing to mixed boundary conditions in the next Section.

\subsubsection*{Odd bulk dimensions}\label{Oddbulkdimensions}

By similar methods as those for the even case, we have for $d+1$ odd dimensions,
\begin{equation}
\label{eq:Cimpar}
\begin{aligned}
    K_D(t,\Omega|X')& := D(\Delta_+) \lim_{\rho \to \pi/2}(\cos \rho\  \sigma)^{\Delta_+-d}\log (\sigma \cos \rho)\ \Theta(\sigma-1)\\
    D(\Delta_+) &:= \frac{(-1)^{(d-2)/2}2^{\Delta_+-d}\Gamma(\Delta_+-\frac{d}{2}+1)}{\pi^{(d+2)/2}\Gamma(\Delta_+-d+1)} \ .
\end{aligned}
\end{equation}
We see that the main difference with $d+1$ even dimensions is the appearence of a $\log(\sigma \cos \rho)$ factor. It was shown in \cite{Hamilton:2006az} that K transforms covariantly anyway. For the same reasons as in the even case, when integrating $K \phi$ there will be a divergence at the endpoints of the range of integration for $\Delta_+<d-1$ regardless of the $\log(x)$ factor (that contributes with a divergence bounded from above by $1/x^\epsilon$ for an arbitrarily small $\epsilon>0$ when $x\rightarrow 0$, so it is not relevant). In the particular case of $d=2$ this problem is not present with Dirichlet boundary conditions since $\Delta_+ \geq 1$, unless we consider $\Delta_+=1$ but this is a more subtle scenario since the solutions of the Klein-Gordon equation degenerate and requires further analysis which we will not perform.  In $d=2$ dimensions, then, the problem of extending the kernel $K$ will arise for Neumann boundary conditions, where after the replacement $\nu\rightarrow-\nu$, we will get $K\sim x^{-1-\nu}\log(x)$ for small $x=\sigma \cos\rho$. This behavior is clearly divergent in the window of interest $0<\nu< 1$. In the rest of the section we explain how to extend $K$ appropriately.

\subsubsection*{Dirichlet boundary condition extended to $\Delta_+ < d-1$}

As we discussed before, the global AdS kernel with Dirichlet boundary conditions is not suited for the case $\Delta_+<2$ for $d=3$, since  a divergence appears in (\ref{eq:Kglobaleven}). Here we will show how to extend the original kernel $K$ of \cite{Hamilton:2006az} of global AdS$_{d+1}$ to the lowest possible values of the dimension, $d/2<\Delta_+<d-1$. This is the same as saying $0<\nu<d/2-1$. We will later use such extension to construct the kernel with Neumann boundary conditions easily. 

We first consider the simplest case $d=3$. The idea is to make an analytic continuation in $\nu$ of the bulk field. Since the bulk field is a solution of the Klein-Gordon equation and its dependence in $\nu$ is analytic, the expression $\int K \phi$ should be analytic in $\nu$. 

We start by defining the bulk field at the origin as a function of $\nu$,
$$ \tilde{\Phi}(\nu):=\Phi_D(\vec{0})$$
as well as
$$f(\nu):=C(\Delta_+)\int d\Omega \sqrt{-h} \int_{-\pi/2}^{\pi/2}dt \phi(t,\Omega) (\cos t)^{-\Delta_-}. $$ 
We know that for $\nu>1/2$ both expressions coincide, $\tilde\Phi=f$. We would like to find an analytic extension of $f$, that we call $g$. We take $f$ and add something  that vanishes for $\nu>1/2$ to cancel the divergence. It is clear that $\phi(t, \Omega)\ \left(\frac{\cos t }{\frac{\pi}{2}-t}\right)^{-\Delta_-}$ is analytic around $t=\frac{\pi}{2}$, so let 
$$\phi(t, \Omega)\ \left(\frac{\cos t }{\frac{\pi}{2}-t}\right)^{-\Delta_-}=\sum_{n=0}^{\infty}b_n^+\ (\pi /2-t)^n$$
be its Taylor series around that point. Analogously, let $\sum_{n=0}^{\infty}b_n^-\ (t+\pi /2)^n$ be the Taylor series of $\phi(t, \Omega)\ \left(\frac{\cos t }{t+\frac{\pi}{2}}\right)^{-\Delta_-}$ around $t=-\frac{\pi}{2}$. . Note that $b_0^\pm=\phi(\pm\pi/2,\Omega)$. Then, we can define
\begin{equation}
\label{eq:extension}
g(\nu):=C(\Delta_+) \int d\Omega \sqrt{-h} \lim_{\epsilon \to 0} \left( (b_0^+ +b_0^-) \frac{\epsilon^{-\Delta_- +1}  }{-\Delta_- +1}+ \int_{-\pi /2+\epsilon}^{\pi /2-\epsilon} dt \phi(t, \Omega)\left( \cos t \right)^{-\Delta_-}\right)\ .
\end{equation}

In order to prove that $g$ extends $f$, let us start by noticing that $g$ is well defined for $\nu>0$, since
\begin{equation}
\begin{aligned}
g(\nu)&=C(\Delta_+) \int d\Omega \sqrt{-h} \lim_{\epsilon \to 0} \Bigg(( (b_0^+ +b_0^-) \frac{\epsilon^{-\Delta_- +1}  }{-\Delta_- +1}+ \int_{0}^{\pi /2-\epsilon} (\pi /2-t)^{-\Delta_-} \sum_{n=0}^{\infty}b_n^+\ (\pi /2-t)^n dt \\
&+\int_{-\pi /2+\epsilon}^{0} (t+\pi /2)^{-\Delta_-} \sum_{n=0}^{\infty}b_n^-\ (t+\pi /2)^n dt \Bigg)\\
&=C(\Delta_+) \int d\Omega \sqrt{-h} \lim_{\epsilon \to 0} \Bigg(( (b_0^+ +b_0^-) \frac{\epsilon^{-\Delta_- +1}  }{-\Delta_- +1}- \sum_{n=0}^{\infty}b_n^+\ \frac{1}{n - \Delta_- +1}((\pi /2-t)^{n - \Delta_- +1})\bigg\rvert_{0}^{\pi /2-\epsilon} \\
&+\sum_{n=0}^{\infty}b_n^-\ \frac{1}{n - \Delta_- +1}((t+\pi /2)^{n - \Delta_- +1})\bigg\rvert_{-\pi /2+\epsilon}^{0} \Bigg) \\
&=C(\Delta_+) \int d\Omega \sqrt{-h} \lim_{\epsilon \to 0} \Bigg(( (b_0^+ +b_0^-) \frac{\epsilon^{-\Delta_- +1}  }{-\Delta_- +1}- \sum_{n=0}^{\infty}b_n^+\ \frac{1}{n - \Delta_- +1}((\epsilon)^{n - \Delta_- +1}-(\pi /2)^{n - \Delta_- +1}) \\
&+\sum_{n=0}^{\infty}b_n^-\ \frac{1}{n - \Delta_- +1}((\pi /2)^{n - \Delta_- +1}-(\epsilon)^{n - \Delta_- +1}) \Bigg)  \\
&=C(\Delta_+) \int d\Omega \sqrt{-h} \ \Bigg(\sum_{n=0}^{\infty}b_n^+\ \frac{1}{n - \Delta_- +1}(\pi /2)^{n - \Delta_- +1}+\sum_{n=0}^{\infty}b_n^-\ \frac{1}{n - \Delta_- +1}(\pi /2)^{n - \Delta_- +1} \Bigg) < \infty \ ,
\end{aligned}
\end{equation}
where in order to obtain the last line we used that if $d=3$, then $n-\Delta_- +1>0$ if $n>0$, which implies that the terms with  $\epsilon$ go to zero. Also we need to prove that $g(\nu)=f(\nu)$ for $\nu>\frac{d}{2}-1$, which is a direct consequence of the fact that in this case $-\Delta_-+1>0$ and then the regulating term in  (\ref{eq:extension}) goes to 0 and the integral is just $f(\nu)$ which converges.

It only remains to see that $g(\nu)=\widetilde{\Phi}(\nu)$ if $\nu>0$. To achieve this let us note that $g$, as $f$, is analytic in $\nu$, because this could only fail if $\nu$ is such that $-\Delta_- +1=0$. However, this zero is cancelled by the pole in the Gamma function coming from $C(\Delta_+)$ (see (\ref{eq:Ceven})). Therefore, we have two analytic functions, $g(\nu)$ and $\widetilde\Phi(\nu)$, with the same connected domain   and that agree in an interval ($\nu>1/2$), thus by the identity theorem we can conclude that $\widetilde{\Phi}(\nu)=g(\nu)$ for $\nu>0$. 

Note that what we have here is essencially the same procedure as what is done to extend the Gamma function integral representation to the whole complex plane. Actually this is the seed to treat more general cases and extend distributions that depend on some parameter, as nicely explained in \cite{Gelfand}. Basically, we are performing an analytic continuation in $\lambda$ of the distribution $P(x) x_+^\lambda$ with $P(x)$ some nice function. More precisely, we are taking $(\cos t)^{-\Delta_-}= (\frac{\cos t}{\pi/2-t})^{-\Delta_-} (\pi/2-t)^{-\Delta_-}$ and then $x=\pi/2-t$, $\lambda=-\Delta_-$ and $P=(\frac{\cos t}{\pi/2-t})^{-\Delta_-}$. The Taylor expansion of $P$ gives a sum of distributions of the form $x_+^\lambda$, each with different $\lambda$. The analytic continuation allows to extend to arbitrary negative $\lambda$ (but with $\lambda$ non negative integer). The factor $\Gamma(\Delta_+-d+1)=\Gamma(-\Delta_-+1)$ in the denominator of $C(\Delta_+)$ actually allows to consider negative integer $\lambda=-\Delta_-$, since it cancels the divergence.     

\begin{figure}[h!] 
\centering
    \includegraphics[width=.5\linewidth]{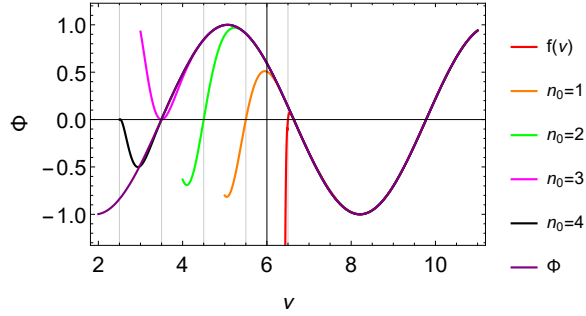} \\
  \caption{We show one mode of $\widetilde\Phi$ as a function of $\nu$ (violet) and the regularized expression of $\int K \phi$ given by (\ref{eq:genericextension}) for $n_0=1,2,3,4$ and $d=15$. The gray vertical lines depict the place where the domain of each curve ends. It is evident that with each additional regulating term, the kernel $K$ can be extended one negative unit in $\nu$. The graphs are truncated because we are using a non-zero regulator $\epsilon=0.01$.  } {\label{fig:regK}}
\end{figure}

From \cite{Gelfand}, we learn that for the generic case $d\geq 3$ higher order terms in  the Taylor expansion need to be included. Calling $n_0$ the number of regulating terms,
\begin{equation}
\label{eq:genericextension}
g(\nu):=C(\Delta_+) \int d\Omega \sqrt{-h} \lim_{\epsilon \to 0} \left(\sum_{j=0}^{n_0-1} (b_j^+ +b_j^-) \frac{\epsilon^{-\Delta_- +1+j}  }{-\Delta_- +1+j}+ \int_{-\pi /2+\epsilon}^{\pi /2-\epsilon} dt \phi(t, \Omega)\left( \cos t \right)^{-\Delta_-}\right)\ .
\end{equation}
extends the original HKLL expression $\int K \phi$ for arbitry values of $\nu$. Note that we can read off the extended kernel $K$ and each regulating term can be interpreted as a derivative of a delta function, implying that $K$ is not really an integral kernel. Figure \ref{fig:regK} compares $g$ and $\widetilde\Phi$ for different number of regulating terms.

\subsection{HKLL adapted to Robin boundary conditions}

\subsubsection*{Even bulk dimensions}

First of all, in light of the previous Section, we should now take the following limit to get the boundary field:
\begin{equation}
\phi:=\lim_{\rho\rightarrow \pi/2} \frac{\Phi_N}{(\cos\rho)^{\Delta_-}}
\end{equation}

We begin with the simplest case,  $1+1$-dimensional AdS and Neumann boundary condition. Instead of going through all the labor of \cite{Hamilton:2005ju} we can just replace $\Delta_+$ with $\Delta_-$ in (\ref{eq:Kglobal1+1})\footnote{The reader may wonder if this is an admissible step, since as discussed previously, the fact that the frequencies were equispaced was crucial in the construction of K using a mode decomposition. However, as also showed in \cite{Hamilton:2006az}, one can alternatively construct a spacelike-supported Green function and read from it $K$. Such Green function can be obtained directly from the Klein-Gordon equation with a Dirac delta source, written in terms of $\sigma$. Since the Neumann boundary condition is AdS invariant, the AdS invariant Green function we need is the same as that of \cite{Hamilton:2006az} and everything goes through. Having said this, we will comment in more detail the relation between Neumann and Dirichlet solutions of the Klein-Gordon equation below, taking into account their different frequencies.}. This is not a convergent expression now, for the exponent is $-\Delta_+$ which can get below $-1$ if $\nu>1/2$. We can start by regularizing the integral, just to give another point of view with a more hands-on feeling, before implementing an analytic continuation argument that anyway is actually equivalent. The integral is divergent at the boundaries of the integration domain. So  we integrate up to these values $\mp \epsilon$  and add a term that takes into account this regularization, i.e. it should have the values of the integrand for $\epsilon\rightarrow 0$ and should vanish for $\nu<1/2$. The final result is

\begin{equation}
\label{eq:kernel1+1}
\begin{aligned}
\Phi_{N} (t', \rho')= &2C(\Delta_-) \lim_{\epsilon \to 0} \Bigg[ \left( \phi(\pi /2-\rho'+t')+\phi(-\pi /2+\rho'+t')\right)\ \frac{\epsilon^{\Delta_-} }{\Delta_-}\\
& +\int_{-\pi /2+\rho'+t'+\epsilon}^{\pi /2-\rho'+t'-\epsilon} dt \phi(t) \left(\lim_{\rho \to \pi /2}\ \sigma\ \cos \rho \right)^{-\Delta_+} \Bigg]\ .
\end{aligned}
\end{equation}
It is straightforward that the $\epsilon^{\Delta_-}$ term is zero for $\nu<1/2$. We can check it is finite for $\nu>1/2$, for example at the origin of AdS$_2$ we have:
\begin{equation*}
\begin{aligned}
 \Phi_{N} (\vec{0})&= 2C(\Delta_-) \lim_{\epsilon \to 0} \Bigg[ \left( \phi(\pi /2)+\phi(-\pi /2)\right)\ \frac{\epsilon^{\Delta_-} }{\Delta_-} +\int_{-\pi /2+\epsilon}^{\pi /2-\epsilon } dt \phi(t) \cos^{-\Delta_+}t \Bigg]\\
& = 2C(\Delta_-) \lim_{\epsilon \to 0} \Bigg[ \int_{-\pi /2+\epsilon}^{\pi /2-\epsilon } dt \frac{d}{dt} \left( \phi(t)\ \sin t\ \frac{\cos^{\Delta_-}t }{\Delta_-} \right)+\int_{-\pi /2+\epsilon}^{\pi /2-\epsilon } dt \phi(t)\ \cos^{-\Delta_+}t  \Bigg] \\
& = \frac{2C(\Delta_-)}{\Delta_-} \lim_{\epsilon \to 0} \Bigg[ \int_{-\pi /2+\epsilon}^{\pi /2-\epsilon } dt \ \phi(t) \left( \cos t \right)^{1+\Delta_-}\left(\Delta_- + 1 \right)+ {\phi}'(t)\ \sin t\ \cos^{\Delta_-}t   \Bigg] \\
& = \frac{2C(\Delta_-)}{\Delta_-} \Bigg[ \int_{-\pi /2}^{\pi /2 } \ \phi(t)\left( \cos t \right)^{2-\Delta_+}\left(\Delta_- + 1 \right))+{\phi}'(t)\ \sin t\ \cos^{\Delta_-} t\  dt\Bigg] <\infty\ ,
\end{aligned}
\end{equation*}
Again, it is interesting to consider the particular case of the massless scalar field:
\begin{equation}
\label{eq:kernelm0}
 \Phi^{M=0}_{N} (t', \rho')= \frac{ \phi(\pi /2-\rho'+t')+\phi(-\pi /2+\rho'+t')}{2}\ .
\end{equation}
In this way we see that, for Neumann boundary conditions, a massless scalar field can be expressed at any point in AdS$_2$ as a function of its boundary value at the two points where the null geodesics meet the boundary. This is in clear contrast with the result obtained for the Dirichlet case (\ref{eq:sinmasa}), where every boundary point spacelike separated from the bulk point contributes equally.

At first sight it may be surprising that the  kernel found for Dirichlet boundary conditions in $d=3$ (\ref{eq:extension}) is almost identical to the one found for Neumann boundary conditions in $d=1$ (\ref{eq:kernel1+1}),  replacing $\Delta_+$ by $\Delta_-$. In fact, in the following we will see that if $\Phi_{N}$ is the scalar field solution with Neumman boundary condition, we have
\begin{equation}
\label{eq:prop}
\widetilde{\Phi}_{N}(\nu)=\widetilde{\Phi}_{D}(-\nu)\ ,
\end{equation}
where $0<\nu<1$ and $\widetilde{\Phi}_{D}$ is the solution with Dirichlet boundary condition as a function of $\nu$.

Therefore, once this is proved, the kernel of Neumann boundary condition is immediately constructed from the one with Dirichlet boundary condition. The only work to do is to extend $g(\nu)$, found in  the previous Subsection, to the range $-1<\nu$. This can be done by adding one more term to the regulator. In the case of even AdS$_{d+1}$ this corresponds to
\begin{equation}
\label{eq:extension3}
g(\nu)=C(\Delta_+)\int d\Omega \sqrt{-h} \lim_{\epsilon \to 0} \left(\sum_{j=0}^{n_0} (b_j^+ +b_j^-) \frac{\epsilon^{-\Delta_- +1+j}  }{-\Delta_- +1+j}+ \int_{t^-+\epsilon}^{t^+-\epsilon} dt \phi(t, \Omega)\left(\lim_{\rho \to \pi/2}\sigma \cos \rho \right)^{-\Delta_-}\right)\ .
\end{equation}
In order to prove (\ref{eq:prop}), remember that the radial solution to the scalar field with Dirichlet boundary condition is $\Phi_3$  while for Neumann boundary condition is $\Phi_4$  (see (\ref{Phi3Phi4})), each with their corresponding frequencies \cite{Balasubramanian:1998sn} $\omega_{(D,N)}=\pm(\Delta_{(+,-)}+l+2n)$, with $n\in\mathbb{N}$ . It is straightforward to check that both solutions are related by the change $\nu\mapsto -\nu$. Then, by replacing (\ref{eq:extension3}) in (\ref{eq:prop}) we get
\begin{equation}
\label{eq:holoneumann}
\begin{aligned}
\Phi_{N}(t',\Omega', \rho')&=C(\Delta_-)\int d\Omega \sqrt{-h} \lim_{\epsilon \to 0} \bigg(\sum_{j=0}^{n_0+1}( b_j^+ +b_j^-) \frac{\epsilon^{-\Delta_+ +1+j}  }{-\Delta_+ +1+j}\\
&+ \int_{t^-+\epsilon}^{t^+-\epsilon} dt \phi(t, \Omega)\left(\lim_{\rho \to \pi/2}\sigma \cos \rho \right)^{-\Delta_+}\bigg)\ ,
\end{aligned}
\end{equation}
We should note that by obtaining the Neumann expression from an analytic continuation of the Dirichlet kernel,  $K_N$ inherits all the nice properties of $K_D$. In particular, the spacelike support. And again, $K_N$ contains derivatives of the delta function.

Finally, in order  to obtain the holographic expression of the bulk field with Robin boundary condition, $\Phi_{R}$, we note that 
\begin{equation}
\label{eq:otrarobin}
	\Phi_{R}(t', \Omega', \rho')= \cos \gamma\ \Phi_D(t', \Omega', \rho')+\sin \gamma\ \Phi_{N}(t', \Omega', \rho')\ ,
\end{equation}
where  $\Phi_D$ and $\Phi_{N}$ have the radial solutions $\Phi_3$ and $\Phi_4$ respectively, but with the frequencies given by (\ref{Robinfreq}). Note that by direct inspection the Dirichlet and Neumann kernels do not depend on these frequencies, contrary to what is perhaps usually claimed based on the original proposals \cite{Hamilton:2005ju,Hamilton:2006az}. Because of this there is no obstacle in using the Robin frequencies in (\ref{eq:otrarobin}) which only enter through the value of the field at the boundary $\phi$ and not in the corresponing $K$'s. Then,
\begin{equation}
K_R=\cos\gamma\ K_D+\sin\gamma\ K_N
\end{equation}

\subsubsection*{Odd bulk dimensions}

In this case the procedure can be repeated as in the even dimensional case. The only difference is that there are two relevant Taylor series, the one of $\phi(t, \Omega)\ \left(\frac{\cos t }{\frac{\pi}{2}-t}\right)^{-\Delta_-}$ and that of $\phi(t, \Omega)\ \left(\frac{\cos t }{\frac{\pi}{2}-t}\right)^{-\Delta_-}\log\left(\frac{\cos t}{\pi/2-t}\right)$.  We present  the expression of $K$ in the case of Dirichlet boundary conditions with $d=2$ for simplicity and because the extension to higher dimensions and Neumann boundary conditions follows the same lines as before:

\begin{equation}
\begin{aligned}
\Phi_D(\vec{0})=& D( \Delta_+) \int d\Omega \sqrt{-h} \lim_{\epsilon \to 0} \Bigg[ \left( \phi(\pi /2, \Omega)+\phi(-\pi /2, \Omega)\right)\ \frac{\epsilon^{-\Delta_-+1} }{-\Delta_-+1} \log(\epsilon) \\
&+\int_{-\pi /2+\epsilon}^{\pi /2-\epsilon } dt \phi(t , \Omega)\left( \cos t \right)^{-\Delta_-} \log(\cos t) \Bigg]\ .
\end{aligned}
\end{equation} 

\section{Boundary-to-bulk map on AdS causal wedges}\label{HKLLwedges}

\subsection{The map $K$ with Dirichlet boundary conditions revisited}

Let us begin discussing the physics in the Poincaré patch. In the original references \cite{Hamilton:2005ju,Hamilton:2006az}  the smearing function for the Poincaré patch was constructed from the one on global AdS. We stick to their notation.  The final result in $1+1$ dimensions reads
\begin{equation}
K(T|T',Z')=\frac{2^{\Delta_+ -1}\Gamma(\Delta_+ +1/2)}{\sqrt{\pi} \Gamma(\Delta_+)} \lim_{Z\to 0} \left(Z \sigma(T,Z|T',Z')\right)^{\Delta_+-1} \Theta(\sigma-1) \ ,
\end{equation}
while for $d+1$ even dimensions ($d>1$),
\begin{equation}
K(T,\vec{X}|P)=\frac{(-1)^{(d-1)/2}2^{\Delta_+-d+1}\Gamma(\Delta_+-d/2+1)}{\pi^{d/2}\Gamma(\Delta_+-d+1)}\ \lim_{Z \to 0}\left(\sigma(T,\vec{ X},Z|P) Z \right)^{\Delta_+-d}\ .
\end{equation}
The case of odd bulk dimensions is similar but with a logarithmic dependence, just as in the global case. We should say that these maps are meant to act not in the global AdS boundary field $\phi$  defined in the previous section, but on a rescaled one defined by
\begin{equation}
\text{Poincaré patch:}\qquad \phi:=\lim_{Z\rightarrow 0} Z^{-\Delta_+}\Phi
\end{equation}
where we are abusing notation and in this section still calling $\phi$ this slightly new boundary field. 

An extension of these expressions to accomodate a dimension down to the unitary bound can be performed following the same lines as in the global AdS case. However, this procedure does not reflect the simple nature of $K$ on AdS causal wedges, where a Fourier transform can be used and simplifies greatly all the computations. The Poincaré patch is a special case of a causal wedge, so we study it directly in momentum space. Moreover, let us concentrate in the $2+1$ dimensional case, since higher dimensions can be incorporated easily.

\subsubsection*{Poincaré patch in $2+1$ dimensions}

We will proceed following \cite{Morrison} and go beyond to show that the kernel $K$ has spacelike support and is in fact real, different from that of \cite{Hamilton:2006az} which contains a logarithm and is non-real in the coordinate representation (below we also compare to the momentum representation in \cite{Hamilton:2006az}). Let us first write the metric as
\begin{equation}
ds^2=\frac{-dT^2+dZ^2+dX^2}{Z^2}
\end{equation}
and expand the solutions of the Klein-Gordon equation as $\Phi\propto\int d^2k e^{-i\omega T+i k_X X} V_{k}(Z)$. Then it follows that for Dirichlet boundary conditions at $Z=0$ we have,
\begin{equation}
V_{k}(Z)=Z J_{\nu}(\sqrt{-k^2}Z),\qquad k^2=-\omega^2+k_X^2.
\end{equation}
Since the small $Z$ limit is $V_k(z)\sim 2^{-\nu}(-k^2)^{\nu/2} Z^{\Delta_+} /\Gamma(\Delta_+)$, the kernel is
\begin{equation}\label{DirichletKPoincare}
K(T,X|T',X',Z')= 2^\nu \Gamma(\Delta_+)  \int_{\mathbb{R}^2} \frac{d\omega\, dk_X}{(2\pi)^2} e^{-i \omega (T'-T) + i k_X (X'-X) } (\omega^2-k_X^2)^{-\nu/2} V_{k}(Z')
\end{equation} 
This is slightly different from \cite{Hamilton:2006az}, but the difference is crucial. In that reference the integral is in the range $|\omega|>|k_X|$, however the expression $\int K \phi$ using (\ref{DirichletKPoincare}) correctly reproduces the bulk field if the Fourier transform of $\phi$ has timelike support. In other words, there is no need to enforce the modes of $K$ to be timelike. The reader may wonder about the ambiguity in the integrand in the range $|\omega|\leq |k_X|$, however the integrand is analytic in $k^2$ (see below). $K$ is also manifestly real. 

The kernel is actually spacelike supported: assume we have a non-spacelike separation\footnote{The invariant distance in Poincaré coordinates is $\sigma=\frac{\Delta X^2+\Delta Z^2-\Delta T^2}{2 Z Z'}$.} $\sigma\leq 1$, and since we have one point at the boundary, we set $Z=0$ and then we look at $\sigma Z$ in this limit: 
$$ \Delta X^2+\ Z'^2-\Delta T^2 \leq 0 \Rightarrow  Z'^2<\Delta T^2 \quad\text{for all}\quad \Delta X^2\neq 0 .$$ 
Then, when $\Delta T+ Z'<0 $ taking a contour of integration in the complex $\omega$ upper-half plane allows to perform the integral in $\omega$. It is important that $V_k(Z)$ is analytic in $\omega$ in the whole complex plane, and this accurs thanks to the factor $(\omega^2-k_X^2)^{-\nu/2}$ which cancels the non-analytic factor of $J_\nu(\sqrt{\omega^2-k_X^2} Z)$. After noting this, the procedure is standard and the integral on the curve such that  $\omega=R e^{i\theta}$, with $\theta\in(0,\pi)$ and fixed large $R$, goes to zero\footnote{We are using the asymptotic expression $J_\nu(z)\sim  z^{-1/2}\cos(z-\frac{(\nu+1/2) \pi}{2}) $ for large $|z|$ \cite{DLMF}.}. In the case that  $\Delta T>Z'$ the contour of integration needs to be taken in the lower-half plane. If one assumed spacelike separation, then the contours for large $|\omega|$ would not go to zero and the previous argument would not work.  Note that in \cite{Hamilton:2006fh} the spacelike support property was shown to hold but through a Wick rotation in the boundary spatial coordinate. We claim this somewhat strange procedure is not necessary if $K$ is defined using all momenta and taking into account that the boundary field has only timelike support in momentum space. 

\subsubsection*{1+1  Rindler wedge}

We include this case just as a warm up for the $2+1$ dimensional one. Here we depart from the original work \cite{Hamilton:2005ju} since there an effort is made towards constructing $K$ so that it works even when the bulk point is behind the horizon, i.e. outside the causal wedge. Instead, we are interested in a kernel $K$ defined only inside the wedge, and we follow \cite{Morrison} but adapted to one less spacelike dimension. First consider the AdS-Rindler metric
\begin{equation}
ds^2=\frac{1}{z^2}\left[-(1- z^2)d\eta^2 + (1- z^2)^{-1}dz^2 \right]
\end{equation}
where $z\in(0,1]$. The boundary field is again different from that of the Poincaré case:
\begin{equation}
\text{AdS Rindler:}\qquad \phi:=\lim_{z\rightarrow 0} z^{-\Delta_+} \Phi
\end{equation}
The kernel reads 
\begin{equation}
K(\eta| \eta', z')= \int \frac{d\omega}{2\pi} e^{-i \omega (\eta-\eta') } v_\omega(z')
\end{equation}  
with $v_\omega(z)$ the Dirichlet mode of the Klein-Gordon equation with frequency $\omega$:
\begin{equation}
v_\omega(z)=z^{\Delta_+} (1- z^2)^{-i\omega/2} {}_2F_1(\frac{\Delta_+-i\omega}{2},\frac{1+\Delta_+-i\omega}{2};\frac{1}{2}+\Delta_+; z^2)
\end{equation}
This kernel is not a priori well-defined, since the modes $v_\omega$ grow like a power of $\omega$. However, taking into account the discussion in the Introduction, we will see in the more interesting case of $2+1$ dimensions  that this $K$ is actually a good map between correlators (as shown in \cite{Morrison}) and even more that is spacelike supported.
 
\subsubsection*{2+1 Rinder wedge}

As discussed previously, the kernel $K$ for the Rindler wedge was properly treated by Morrison in \cite{Morrison}. Consider the metric 
\begin{equation}
ds^2=\frac{1}{z^2}\left[-(1-z^2)d\eta^2 + (1- z^2)^{-1}dz^2 +d\chi^2 \right]
\end{equation}
where again $z\in(0,1]$. The kernel then reads 
\begin{equation}\label{DirichletKRindler}
K(y|x',z')= \int \frac{d^2k}{(2\pi)^2} e^{i k\cdot(x'-y) } V_{k}(z')
\end{equation}  
where $k\cdot x=-\omega \eta + k^1 \chi $, and with $V_k(z)$ the Dirichlet mode of the Klein-Gordon equation with two-dimensional momentum $k$:
\begin{equation}
V_k(z)=z^{\Delta_+} (1-z^2)^{-i\omega/2} {}_2F_1(\frac{\Delta_+-i\omega+ik^1}{2},\frac{\Delta_+-i\omega-ik^1}{2};\Delta_+; z^2)
\end{equation}
The cautionary comment in the $1+1$ case regarding the power-law growth of the modes is in this case worth revisiting. For timelike and null momenta the conslusion remains. For spacelike momentum the growth  is exponential. It was an important observation of \cite{Morrison} that this is not a problem if one uses boundary test functions constructed from compactly supported bulk test functions as $f_F=\int_{\text{supp}(F)} K F$, as discussed in the Introduction. 

Let us show that this $K$ is actually spacelike supported. First of all, the invariant distance in these coordinates is given by
\begin{equation}\label{sigmaRindler}
\sigma=\frac{\cosh\Delta\chi-\sqrt{(1-z^2)(1-z'^2)} \cosh\Delta\eta}{z z'}
\end{equation}  
Spacelike separation between a bulk point and a boundary point means that $\sigma z >0 $ in the limit $z\rightarrow 0$, while timelike separation means $\sigma z <0$. It is convenient to introduce an alternative radial coordinate: 
$$ r:=\tanh^{-1} z ,\qquad r\in (0,\infty) $$
then timelike separation means 
\begin{equation}\label{spacelikesupport}
\cosh r \cosh\Delta\chi <\cosh\Delta\eta
\end{equation}
which implies that a necessary condition for timelike support is $r<|\Delta\eta|$. Let us now write $K$ as
\begin{equation}
K(y|x',z')= \int \frac{dk^1}{(2\pi)} e^{i k^1 \Delta\chi} \int \frac{d\omega}{(2\pi)} e^{-i\omega \Delta\eta} V_{k}(z')
\end{equation}  
and perform a complex integral similar to that in the Poincaré patch. The integrand is analytic in $\omega$ and then we can show the integral is zero by analyzing the behavior for large $|\omega|$, more precisely $|\omega|>>|k^1|,\Delta_+>0$. In this particular limit case \cite{Morrison}, 
\begin{equation}
V_k \sim  |\omega|^{1/2-\Delta} \sqrt{\tanh r} \left( e^{ir\omega} + e^{-ir\omega}\right)
\end{equation}  
where we omitted unimportant factors. Then, the integrand can be bounded as (with $\omega =R e^{i\theta}$)
\begin{equation}
\left| e^{-i\omega \Delta\eta} V_k\right| \leq R^{1/2-\Delta} \left( e^{R\sin\theta(\Delta \eta-r)}+ e^{R\sin\theta(\Delta \eta+r)}\right) \rightarrow 0
\end{equation}
where the limit is obtained in the timelike case $r<|\Delta\eta|$ and choosing the sign of Im $\omega$ appropriately. We conclude then that $K$ is actually spacelike and  lightlike supported. Actually, for $\Delta\chi\neq 0$ it can  only be spacelike supported, by the same arguments. We should stress once again that we have not made a Wick rotation of the boundary spatial coordinate as in \cite{Hamilton:2006az,Hamilton:2006fh}.

\subsection{The extension to Robin boundary conditions}  
\subsubsection*{Poincaré patch}

There are at least two distinct routes to take. First, one can restrict the global K to the Poincaré patch (asuming the boundary test function has support only on the boundary of the Poincaré patch), and there is nothing additional to do but to change coordinates in (\ref{eq:otrarobin}). However if one would like to mimic the analysis in \cite{Hamilton:2005ju,Hamilton:2006az} to go beyond the Poincaré horizon, then the antipodal map is needed and the spacelike support is a required feature, but as remarked in \cite{Hamilton:2006az} the kernel obtained is not spacelike supported. 

The second route is to take advantage of the expression of $K$ for the Dirichlet case as a Fourier integral operator as in (\ref{DirichletKPoincare}). We already showed it is real and spacelike supported.  Then, as already discussed, the Neumann $K$ is obtained by just replacing $\nu\rightarrow -\nu$, and the Robin kernel is the linear combination 
\begin{eqnarray}\label{RobinKPoincare}
K_R(T,X|T',X',Z')&=&\int_{\mathbb{R}^2} \frac{d\omega\, dk_X}{(2\pi)^2} e^{-i \omega (T'-T) + i k_X (X'-X) }
\left[ \cos\gamma\,\,  2^\nu \Gamma(\Delta_+)   (\omega^2-k_X^2)^{-\nu/2} Z' J_\nu(\sqrt{-k^2}Z')\right.\nonumber\\
&+&\left. \sin\gamma\,\, 2^{-\nu} \Gamma(\Delta_-)   (\omega^2-k_X^2)^{\nu/2} Z' J_{-\nu}(\sqrt{-k^2}Z')\right]
\end{eqnarray}
where $k^2=-\omega^2+k_X^2$. 

\subsubsection*{Rindler causal wedge}

The way to extend (\ref{DirichletKRindler}) from $\Delta_+$ to $\Delta_-$ is identical as the previous case. The advantage comes from having expressed $K$ as a Fourier transform of $V_k(z)$. Then, instead of extending $K$ looking at where its singularities are in position space, we can extend $V_k(z)$ and then take its Fourier transform. Since there is nothing that prevents from changing $\nu$ to $-\nu$ in $V_k(z)$ (it is possible to check that the WKB analysis of $V_k(z)$ in \cite{Morrison} remains the same), the kernel with Neumann boundary condition is just as in (\ref{DirichletKRindler}) but with $V_k(z)$ evaluated on $\Delta_-$ instead of $\Delta_+$. Finally, for Robin boundary conditions one takes the same linear combination as in the global and Poincaré cases,
\begin{eqnarray}
K(x|x',z')&=&\int \frac{d^2k}{(2\pi)^2} e^{i k\cdot(x'-x)}\left[ \cos\gamma\,\, z'^{\Delta_+} (1-z'^2)^{-i\omega/2} {}_2F_1(\frac{\Delta_+-i\omega+ik^1}{2},\frac{\Delta_+-i\omega-ik^1}{2};\Delta_+; z'^2) \right.\nonumber\\
&+&\sin\gamma\,\,\left. z'^{\Delta_-} (1-z'^2)^{-i\omega/2} {}_2F_1(\frac{\Delta_--i\omega+ik^1}{2},\frac{\Delta_--i\omega-ik^1}{2};\Delta_-; z'^2)\right]
\end{eqnarray}

\section{Boundary-to-bulk map of correlators along the RG flow}\label{Mapping}

Here we will follow the strategy and conclusions of  \cite{Morrison} and perform some more explicit computations (in particular we show in the following that the sensible assumption of microlocal spectrum condition of \cite{Morrison} is satisfied). We are interested in AdS causal wedges, and in those the  kernel K can be cast in a very short form by means of the Fourier transform. Also, the Fourier transform is crucial to understand the way we can use $K$ to map correlators of the boundary to bulk correlators, and to map bulk test functions to boundary test functions (see the Introduction). For these reasons, we will stick to the Poincaré patch and comment on the generalization to an arbitrary causal wedge and leave the global AdS case aside.

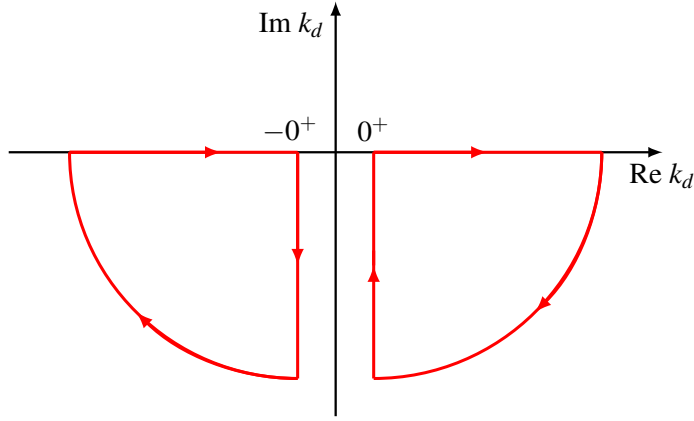
\begin{figure}
	\centering
	\begin{tikzpicture}
	
	\draw[line width=0.3mm,->] (-4.3,1) -- (4.3,1) node[anchor=north] {Re $k_d$};
	\draw[line width=0.3mm,->] (0,-2.5) -- (0,3) node[anchor= north east] {Im $k_d$};
	
	\draw[line width=0.4mm,red] (-3.5,1) -- (-0.5,1) node[anchor=north west] {};
	\draw[line width=0.4mm,red] (-3.5,1) arc (180:270:3);
	\draw[line width=0.4mm,red] (-0.5,1) -- (-0.5,-2) node[anchor=south east] {};
	\draw[line width=0.4mm,red, ->] (-3.5,1) -- (-1.5,1) node[anchor=north west] {};
	\draw[line width=0.4mm,red,->] (-0.5,1) -- (-0.5,-0.5) node[anchor=south east] {};
	\draw[line width=0.4mm,red,->] (-0.5,-2) arc (-90:-135:3);
		
	\draw[line width=0.4mm, red] (0.5,1) -- (3.5,1) node[anchor=north west] {};
	\draw[line width=0.4mm,red] (3.5,1) arc (0:-90:3);
	\draw[line width=0.4mm,red] (0.5,1) -- (0.5,-2) node[anchor=south east] {};
	\draw[line width=0.4mm,red, ->] (0.5,1) -- (2,1) node[anchor=north west] {};
	\draw[line width=0.4mm,red,->] (0.5,-0.5) -- (0.5,-0.5) node[anchor=south east] {};
	\draw[line width=0.4mm,red,->] (3.5,1) arc (0:-45:3);
	
	 \node at (0.5, 1.3)   (a) {$0^+$};
	 \node at (-0.6, 1.3)   (a) {$-0^+$};
	
	\end{tikzpicture}
	\caption{\label{fig:circ}  Integration contour. }
	\label{fig:contour}
\end{figure}

We would like to make explicit how the kernel $K$ found in the previous section takes a correlator of the perturbed CFT in the boundary and gives back a correlator in the bulk with mixed boundary conditions. We will work with the 2-point Schwinger and Wightman functions. In order to proceed, we will first of all make the analytic continuation of the correlator found in \cite{GubserKlebanov} (see also \cite{HartmanRastelli}), in order to get the corresponding Wightman 2-point function in the boundary. When perturbing the CFT by a double trace term of the form $\frac{f}{2}\int\mathcal{O}^2 $, with a relevant single-trace operator of weight $\Delta_-$ , the Schwinger 2-point function reads   
\begin{equation}
s((\tau,\vec{x}); 0)= \int \frac{d^{d-1}k}{(2\pi)^{d-1}} e^{i \vec{k}.\vec{x}} \int_{-\infty}^\infty \frac{dk_d}{2\pi} e^{i k_d \tau} \frac{A_\nu}{k^{2\nu} +f A_\nu} 
\end{equation} 
where $k^2$ denotes the Euclidean squared momentum and 
\begin{equation}
A_\nu=2^{2\nu} \pi^{d/2}\frac{\Gamma(\nu)}{\Gamma(\frac{d}{2}-\nu)} 
\end{equation}
As mentioned, we have to obtain the corresponding Wightman function. The spatial Fourier integral will play no part, so we concentrate on the $dk_d$ integral. Take $\tau=-i(t-i\epsilon)$,
\begin{equation}
\int_{-\infty}^\infty \frac{dk_d}{2\pi} e^{i k_d \tau} \frac{A_\nu}{k^{2\nu} +f A_\nu} = \int_{-\infty}^\infty \frac{dk_d}{2\pi} e^{ k_d t-i k_d \epsilon} \frac{A_\nu}{({\vec{k}}^{2}+k_d^2)^{\nu} +f A_\nu} 
\end{equation} 
We are going to perform the integral going to the complex plane, and we take the branch cut in the negative imaginay axis. Now, the contour of integration we choose is given in Figure \ref{fig:contour}. With the definition $\omega=i k_d$, with $\omega>0$ along the contour, it is easy to see that 
\begin{equation}
\int_{-\infty}^\infty \frac{dk_d}{2\pi} e^{i k_d \tau} \frac{A_\nu}{k^{2\nu} +f A_\nu} =i \int_0^\infty \frac{d\omega}{2\pi} e^{ -i\omega t- \omega \epsilon} A_\nu \left[\frac{1}{({\vec{k}}^{2}-(w-i 0^+)^2)^{\nu} +f A_\nu} -\frac{1}{({\vec{k}}^{2}-(w+i 0^+)^2)^{\nu} +f A_\nu} \right]
\end{equation} 
Now using that 
\[({\vec{k}}^{2}-(\omega\pm i 0^+)^2)^\nu=(|\vec{k}|+\omega\pm i 0^+)^\nu(|\vec{k}|-\omega\mp i 0^+)^\nu= (|\vec{k}|+\omega)^\nu (|\vec{k}|-\omega\mp i 0^+)^\nu \] 
after some straightforward manipulations,
\begin{eqnarray}
\int_{-\infty}^\infty \frac{dk_d}{2\pi} e^{i k_d \tau} \frac{A_\nu}{k^{2\nu} +f A_\nu} &=&i \int_0^\infty \frac{d\omega}{2\pi} e^{ -i\omega t- \omega \epsilon} A_\nu  (|\vec{k}|+\omega)^\nu\nonumber\\
&\times& \left[\frac{({\vec{k}}-\omega-i 0^+)^\nu-({\vec{k}}-\omega+i 0^+)^\nu}{(({\vec{k}}^{2}-(\omega-i 0^+)^2)^{\nu} +f A_\nu)(({\vec{k}}^{2}-(\omega+i 0^+)^2)^{\nu} +f A_\nu)}\right]
\end{eqnarray}
The following result of generalized functions comes in handy \cite{Gelfand}, 
\[ (x + i0^+)^\nu - (x - i0^+)^\nu = 2 i \sin(\pi \nu) \theta(-x) |x|^\nu \]
so taking $x = |\vec{k}|-\omega$ we see that we get the condition of support in the future lightcone, and then we can also take $({\vec{k}}^{2}-(\omega\pm i 0^+)^2)^{\nu}=(|\vec{k}|+\omega)^\nu (\omega-|\vec{k}|)^\nu e^{\mp i\pi\nu}=(-p^2)^\nu e^{\mp i\pi\nu}$, with $p^2=-\omega^2+\vec{k}^2$. Finally,
\begin{eqnarray}
\int_{-\infty}^\infty \frac{dk_d}{2\pi} e^{i k_d \tau} \frac{A_\nu}{k^{2\nu} +f A_\nu} &=&\frac{A_\nu }{\pi} \int_0^\infty \frac{d\omega}{2\pi} e^{ -i\omega (t- i \epsilon)}  \nonumber\\
&\times& \frac{\sin(\pi\nu) (-p^2)^\nu \theta(-p^2)}{(-p^2)^{2\nu} + fA_\nu 2\cos(\pi\nu)(-p^2)^\nu +(fA_\nu)^2  }
\end{eqnarray}
So the  Wightman 2-point function corresponding to the double-trace deformation in the CFT is given by\footnote{This result, when expressed as an integral over a positive mass parameter as is done in the Appendix, is consistent with the Kallen-Lehmann representation found in the Euclidean setting in \cite{Porrati:2016lzr}}
\begin{equation}\label{boundary2point}
\omega_2(x,0)=\frac{A_\nu \sin(\pi\nu)}{\pi} \int \frac{d^dk}{(2\pi)^{d}}e^{i \vec{k}.\vec{x}-i\omega (t-i\epsilon)}
\frac{(-p^2)^\nu \theta(-p^2)\theta(\omega)}{(-p^2)^{2\nu} + fA_\nu 2\cos(\pi\nu)(-p^2)^\nu +(fA_\nu)^2  } 
\end{equation}

Now we are in a position to apply the kernel K twice and show that it correctly gives the Wightman 2-point function of the bulk scalar field with mixed boundary conditions. It is actually a straightforward calculation. Take the Poincaré kernel (\ref{RobinKPoincare}) and the boundary two-point function (\ref{boundary2point}) and write 
\[ \int d^2x_1 d^2x_2 K_R(X_1;x_1)K_R(X_2;x_2) \omega_2(x_1,x_2)\]
These integrals contribute with two delta functions, which can be used to integrate the momenta corresponding to each $K$. The result is then
\begin{eqnarray}
&&Z_1 Z_2 \sin^2\gamma\, \int \frac{d^2k}{(2\pi)^2} e^{i k\cdot (x_1'-x_2')}\theta(-p^2) \theta(\omega)\times\nonumber\\
&& \frac{\left[\cot\gamma\,\, J_\nu(\sqrt{-p^2}Z_1) + (-p^2)^{\nu} J_{-\nu}(\sqrt{-p^2}Z_1) \right]\left[\cot\gamma\,\, J_\nu(\sqrt{-p^2}Z_2) + (-p^2)^{\nu} J_{-\nu}(\sqrt{-p^2}Z_2) \right]}{(-p^2)^{2\nu} + fA_\nu 2\cos(\pi\nu)(-p^2)^\nu +(fA_\nu)^2}
\end{eqnarray}
This is (modulo a normalization constant) the bulk two-point function  with Robin boundary conditions found in \cite{Dappiaggi:2016fwc}, with the identification
\begin{equation}
\cot\gamma = -f A_\nu
\end{equation} 

We now comment on the microlocal analysis of (\ref{boundary2point}) and in particular we would like to show that it does satisfy the criteria of the microlocal spectrum condition (see (\ref{boundaryWF}) below). It is evident from the expression (\ref{boundary2point}) that the Fourier transform has support on the (closure of the) future lightcone. Then, it remains to see where the singular support of $\omega_2$ is. Of course, by Lorentz invariance we expect it to have singularities whenever the geodesic distance from $x_1$ to $x_2$ is zero (since at the origin $\omega_2$ is singular). However, we would like to be sure there are not singular points other than those. In order to study this, we will use results from \cite{Hormander} and \cite{RS} (see also the very nice notes \cite{Brouder:2014hta} for a smooth introduction to wavefront sets). We leave the details to the Appendix, and just state the  final result 
\begin{equation}\label{boundaryWF}
WF(\omega_2)\subseteq \left\{(x_1,k_1;x_2,k_2)\in (T^*\mathbb{R}^{1,1})^2 \setminus\left\{0\right\} \mid |x_1-x_2|^2=0  ,\quad k_1^2\leq 0,\quad k_1\sim -k_2     \right\}
\end{equation}    
namely the boundary two-point function of the deformed theory satisfies the microlocal spectrum condition and this allows to map bulk test functions (smooth of compact support) to admissible boundary test functions by using $K$, as discussed in the Introduction. The result (\ref{boundaryWF}) can be interpreted as a confirmation of the assumption of \cite{Morrison}, as well as a generalization to the entire RG flow.

\section{Conclusions}

We have shown how to adapt the HKLL map to the case where the bulk field satisfies mixed boundary conditions. This, in particular, allows to consider the case where the dimension reaches the unitary bound $\Delta=(d-2)/2$. Along the way we learnt that actually the original and standard construction of the HKLL map for Dirichlet boundary conditions only works for $\Delta_+>d-1$ and so we extended the map to account for $\Delta_+>d/2$. This procedure  turned out to be of great utility to learn how to analytically continue $K$ to lower values of $\nu$ and with the identification of Neumann boundary conditions as an extension of the range of $\nu$, we were able to construct the maps of HKLL adapted to Neumann boundary conditions.  The case of mixed (Robin) boundary conditions was then straightforward to resolve, once we made a few comments on how the Robin frequencies only appear through the boundary field and not through $K_D$ or $K_N$. 

So far we discussed the global AdS spacetime. We then focused on causal wedges, in particular the Poincaré patch and the Rindler wedge. In both cases we showed that the map $K$ is spacelike and real. Even more, there is such map for Robin boundary conditions, suggesting there is a possible bulk reconstruction even from localised regions on a non conformally-invariant boundary theory.  

From the boundary QFT perspective, the mixed boundary conditions on the bulk are known to be captured by the addition of a double-trace perturbation to the CFT. Then, the fact that there is still a way to define $K$ for any Robin boundary condition means that there is a bulk reconstruction along the RG flow. The question is then if the very interesting insights from \cite{Morrison}, discussed at length in the Introduction, still hold when we break the conformal invariance at the boundary. Namely, in order to be consistent, we asked whether the kernel $K$ can still be used to map correlators of the perturbed CFT to correlators of the bulk theory with mixed boundary conditions, in the Poincaré patch. To this end we showed that the boundary QFT 2-point function satisfies the microlocal spectrum condition (\ref{microlocal})  of \cite{Brunetti:1995rf}. With this result at hand we can follow the logic of \cite{Morrison} and claim that $K$ is a good object to act on boundary correlators, despite being exponentially divergent in (spacelike directions of) momentum space. Even more, we explicitly mapped the boundary 2-point function to the bulk, reproducing the already known two-point function with Robin boundary conditions in the Poincaré patch. 

To conclude, we would like to point out first that the present work could be a starting point for  the exploration of the so called  \textit{subregion duality} with broken conformal invariance at the boundary. For instance, it would be very interesting to adress the problem of mapping higher point functions from the boundary to the bulk for non-Gaussian states. Also, it remains to study the interacting bulk theory considering subleading terms in the $1/N$ expansion. This is a difficult task, since it is believed that a tower of fields of higher dimensions is needed in order to reconstruct the interacting bulk field. Finally, the presence of true event horizons in the bulk seems worth considering, without resorting to the unconventional analytic continuation of the boundary spacelike coordinate in \cite{Hamilton:2006az}.

\appendix

\section{Wavefront sets, oscillatory integrals and microlocal spectrum condition} \label{WF}

Let us start this appendix with a short overview on wave of front sets of distributions. This will allow to state in a clear way the microlocal spectrum condition of \cite{Brunetti:1995rf} for a Wightman 2-point function \footnote{We will only refer to the microlocal behavior of the 2-point functions, however the microlocal spectrum condition of \cite{Brunetti:1995rf} is in fact a condition on all the correlators of the theory and permits, roughly speaking, to assure that they can be combined without loosing control on the singular structure of these correlators.}. Finally we prove that this is satisfied for the perturbed CFT Wightman 2-point function. Most of the time we follow \cite{Hormander} and work with $\mathbb{R}^n$, although the theory of wavefront sets is well-suited for smooth manifolds.
 
Let $f\in C^{\infty}(\mathbb{R}^{n})$ be a function and consider $\Phi_{f}:C^{\infty}_{c}(\mathbb{R}^{n})\to \mathbb{R}$ defined by
\begin{equation}\label{def}
\Phi_{f}(h)=\int_{\mathbb{R}^{n}} h(x)f(x)dx.
\end{equation}
Here $C^{\infty}_{c}(\mathbb{R}^{n})$ denotes the set of compactly supported smooth functions. Note that $\Phi_{f}$ is a linear map and so it is in the dual space of $C^{\infty}_{c}(\mathbb{R}^{n})$. The space of such  continuous linear functionals will be denoted by $C'^{\infty}_{c}(\mathbb{R}^{n})$ (continuity is taken under certain topology, see  \cite{Hormander,rudin}). Actually for \eqref{def} to make sense it suffices to require $f\in L^{1}_{Loc}(\mathbb{R}^{n})$, i.e. that it is a locally integrable function.

From the above comments, one can regard $C'^{\infty}_{c}(\mathbb{R}^{n})$ as a space that generalizes functions, called the space of \textit{distributions} with \textit{test functions}  $C^{\infty}_{c}(\mathbb{R}^{n})$. But distributions do not always come from functions. As an example, a Lebesgue measure $\mu$ is  not generally a function but \eqref{def} still works if we change $f(x)dx$ by $d\mu$. The typical example is the Dirac delta measure $\delta$ whose associated $\Phi_{\delta}$ is defined by $\Phi_{\delta}(h)=h(0)$ but there is no smooth function $f$ such that $\Phi_f=\Phi_\delta$.

Different spaces of test functions $C^{\infty}_{c}(\mathbb{R}^{n})\subseteq \mathcal{S}(\mathbb{R}^{n})\subseteq C^{\infty}(\mathbb{R}^{n})$ give place to different distributional spaces by the same construction as above ($\mathcal{S}(\mathbb{R}^{n})$ denotes the space of Schwarz functions), but the inclusions are reversed, $ C'^{\infty}(\mathbb{R}^{n})\subseteq \mathcal{S}'(\mathbb{R}^{n})\subseteq C'^{\infty}_{c}(\mathbb{R}^{n})$ , these spaces are called compact support distributions, tempered distributions and just distributions respectively. 

A few examples are in order: if $f(x)=e^x$, $\Phi_{f}\in C'^{\infty}_{c}(\mathbb{R})$, but $\Phi_{f}$ doesn't belong to $ C'^{\infty}(\mathbb{R})$ or $\mathcal{S}'(\mathbb{R})$. If $g(x)=e^{-x^2}$ then $\Phi_{g}\in \mathcal{S}'(\mathbb{R})\setminus C'^{\infty}(\mathbb{R})$, because \eqref{def} converges for all $h\in  \mathcal{S}(\mathbb{R})$, but diverges for $h(x)=e^{x^2}$. In order to guarantee convergence of \eqref{def} for all $h\in C^{\infty}(\mathbb{R})$, one needs that the \textit{support of the distribution} be compact (see \cite{rudin} for details). Roughly speaking, the support of a distribution is the complement of the set where the distribution is zero. The $\delta$ distribution has support at the origin and the $\theta$ distribution has support in $[0,+\infty]$.

An important object for us is the \textit{singular support} of a distribution, which is formed by those points in which it fails to be smooth. Namely, the complement of those points where there is an open neighborhood where the distribution is of the form $\Phi_f$ for some smooth function $f$. For instance the origin is the singular support of the $\delta$ distribution and of the $\theta$ distribution, and more generally the boundary of a set is the singular support of the corresponding characteristic function. 

Given two distributions $u,v\in C'^{\infty}_{c}(\mathbb{R}^{n})$, if both are smooth, that is $u=\Phi_{f}$ and $v=\Phi_{g}$ with $f$ and $g$ smooth functions, the product $uv$ is the distribution $\Phi_{fg}$. Moreover, if only $v=\Phi_g$, the product is defined by $uv(h)=u(gh)$, using $gh$ as a test function for $u$. But even in the case when neither of the two distributions are smooth, the product might be defined. To do this we need to introduce the concept of \textit{wavefront set} of a distribution.

For tempered distributions the Fourier transform is defined by $\hat{u}(h)=u(\hat{h})$. From this definition, if $u$ is a  smooth compactly supported distribution then $\hat{u}=\Phi_{f}$ where $f(\xi)= u(g_\xi)$ and $g_\xi:x\mapsto e^{-i\xi\cdot x}$ (see \cite{Hormander} or Chapter 2 \cite{vestberg}, for a nice review of this and many of the following statements).  Let us consider the following inequality
\begin{equation}\label{cota}
|\hat{u}(\xi)|\leq C_{n}(1+|\xi|)^{-n}, \hspace{1cm} n\in \mathbb{N}
\end{equation}
If $u$ is a compactly support distribution which satisfies \eqref{cota} for all $n$, then $u$ must be smooth. Even more, since $u$ has compact support then $u$ comes from a compactly supported smooth function. Then for $u\in C'^{\infty}_{c}(\mathbb{R}^{n})$ the directions $\xi$ for which $u$ does not satisfy \eqref{cota} for some $n$ are responsible for $u$ not being smooth.

Given a distribution $u\in C'^{\infty}_{c}(\mathbb{R}^{n})$ its wavefront set is a subset of $\mathbb{R}^{n}\times\mathbb{R}^{n}\setminus \left\{0\right\}$ containing\footnote{To be precise the wavefront set is exactly the set of point $(p,\xi)\in \mathbb{R}^n\times\mathbb{R}^n\setminus\left\{0\right\}$ such that $p$ is in the singular support of $u$ and $\xi$ does not have a conic neighbourhood $V$ such that $\phi u$  satisfies \eqref{cota} in $V$, for all $\phi\in C^{\infty}_{c}(\mathbb{R}^{n})$ with $\phi(p)\neq 0$ }  those points $(p,\xi)$ such that $p$ is in the singular support of $u$ and $\xi$ is a direction such that $\phi u$ doesn't satisfies \eqref{cota}  for all $\phi\in C^{\infty}_{c}(\mathbb{R}^{n})$ with $\phi(p)\neq 0$ (observe that $\phi u$ is a compactly supported distribution). Geometrically the wavefront set of a distribution can be thought as the points and directions in which the distribution fails to be smooth. An enlightening example is the 2-dimensional  step function $h:\mathbb{R}^{2}\to \mathbb{R}$, $h(x,y)=1$ if $x\geq 0$ and zero in another case, whose wavefront set is given by $WF(h)=\{ (0,y,t,0)|y\in \mathbb{R}, t\neq 0 \}$, that is the problematic directions are those perpendicular to the step. 

The most important result about wavefront set pertains the possibility of multiplying two distributions:

\begin{theorem}{
 \cite[Theorem 8.2.10]{Hormander} }
If $u$ and $v$ are distributions and there is no element $(p,\xi)\in WF(u)$ such that $(p,-\xi)\in WF(v)$. Then the previous definition of the product $uv$ can be generalised.
\end{theorem}

For example if we consider the $\delta$ distribution, since it is a smooth compactly supported distribution, $\hat{\delta}=\Phi_{f}$ with $f(\xi)= \delta(e^{-i\xi\cdot x})=1$, and then $\hat{\delta}=1$ so it does not satisfies \eqref{cota} for any direction. By the previous theorem we cannot define $\delta^2$ as proposed in \cite{Hormander}. On the contrary, a step distribution in the $\hat{n}$ direction and a step distribution in another direction can be multiplied.

We are specially interested in a specific kind of distributions commonly called oscillatory integrals, which are obiquitous in QFT. An oscillatory integral is a formal expression 
\begin{equation}
I_{\varphi}[a](x)=\int_{\mathbb{R}^{s}}e^{i\varphi(x,\theta)}a(x,\theta) d\theta
\end{equation} 
for a function on $\mathbb{R}^n$. Here $\varphi$ is a phase function, $a$ is an asymptotic symbol of some order and $\theta \in \mathbb{R}^s$ (so it may not necessarily be related to the $x$ coordinates by a Fourier transform). A phase function is a function which is continuous  and homogeneous for positive scalars in the $\theta$ variable, which is smooth in $\mathbb{R}^{n}\times(\mathbb{R}^{s}\setminus \{0\})$ and whose gradient $(\nabla_{x}\varphi,\nabla_{\theta}\varphi)$ is never zero for $\theta \neq 0$. $a$ is an asymptotic symbol of order $m$ if for each compact $K\subseteq \mathbb{R}^{n}$, there are constants such that 

\begin{equation}\label{cotasimbolo}
|(D^{\alpha}_{x}D^{\beta}_{\theta}a)(x,\theta)|\leq d_{\alpha,\beta,K}(1+|\theta|)^{m-|\beta|}, \text{ for $x\in K$ and $\theta\in \mathbb{R}^{s}$} 
\end{equation} 
Oscillatory integrals can be considered as well-defined distributions. 

An immediate task that follows is to characterize their wavefront sets, namely their singularity structure. To this end let us introduce two manifolds, 
\begin{align*}
 M(\varphi)&=\{(x,\theta)\mathbb{R}^{n}\times(\mathbb{R}^{s}\setminus \{0\}) | (\nabla_{\theta}\varphi)(x,\theta)=0\} \subseteq \mathbb{R}^{n}\times \mathbb{R}^{s}\\
 SP(\varphi)&=\{(x,(\nabla_{x}\varphi)(x,\theta)) | (x,\theta)\in M(\varphi) \}\subseteq \mathbb{R}^{n}\times \mathbb{R}^{n}
\end{align*}
the latter is called the manifold of stationary phase for $\varphi$. The following theorem provides a very useful constraint on the wavefront set of the oscillatory integral,

\begin{theorem}
\label{wfsetasym}{ \cite[Theorem IX.47]{RS} or \cite[Theorem 8.1.9]{Hormander}}
For any phase function $\varphi(x,\theta)$ and asymptotic symbol $a(x,\theta)$, $WF(I_{\varphi}(a))\subseteq SP(\varphi)$.
\end{theorem}

The microlocal spectrum condition, often denoted $\mu$SC, is a statement about the wavefront sets of the Wightman functions of a QFT \cite{Brunetti:1995rf}. We are just going to express it for the 2-point function $\omega_2$,
\begin{equation}\label{microlocal}
WF(\omega_2)\subseteq \left\{(x_1,k_1;x_2,k_2)\in (T^*\mathbb{R}^{1,1})^2 \setminus\left\{0\right\} \mid |x_1-x_2|^2=0  ,\quad k_1^2\leq 0,\quad k_1\sim -k_2     \right\}
\end{equation}   
where $k_1\sim -k_2$ means the parallel transport of $k_1\in T^*_{x_1}(\mathbb{R}^{1,1})$ from $x_1$ to $x_2$  through a null geodesic coincides with $-k_2$. Roughly speaking, this condition guarantees that the singularity structure of the 2-point function is at most as bad as lying in the lightcones of fixed $x_1$ and with future pointing causal momenta $k_1$.

Now we are ready to turn to the task of proving that the 2-point function of the perturbed CFT \eqref{boundary2point} satisfies $\mu$SC (\ref{microlocal}). We can start rewriting the appropriate integration limits in order to eliminate the factor $\theta(-p^2)\theta(\omega)$  and then apply the change of variables $m^2=\omega^2-|\vec{k}|^2=-p^2$,
\begin{align*}
\omega_2(x,0)=\frac{A_\nu \sin(\pi\nu)}{\pi} \int \frac{d^{d-1}k}{(2\pi)^{d}}e^{i \vec{k}.\vec{x}}\int_{0}^{\infty}dm
\frac{e^{-i\sqrt{m^2+\vec{k}^2}t}}{\sqrt{m^2+\vec{k}^2}}\frac{m^{2\nu+1}}{m^{4\nu} + fA_\nu 2\cos(\pi\nu) m^{2\nu} +(fA_\nu)^2}\\
= \frac{A_\nu \sin(\pi\nu)}{\pi}\int_{0}^{\infty}dm^2
\frac{m^{2\nu}}{m^{4\nu} + fA_\nu 2\cos(\pi\nu) m^{2\nu} +(fA_\nu)^2} \Delta_{+}(x,m^2)\hspace{3cm}
\end{align*}
where $\Delta_{+}(x,m^2)$ is proportional to the free massive Klein-Gordon 2-point function in Minkowski space. Leaving aside multiplicative factors, it is an oscillatory integral with phase function 
\begin{equation}
\varphi(x,\vec{k})=-t|\vec{k}|+\vec{x}\cdot\vec{k} \in \mathbb{R}^{4}\times \mathbb{R}^{3}
\end{equation} 
and 
\begin{equation}
a(x,\vec{k},m^2)=\frac{e^{-it[\sqrt{m^2+\vec{k}^2}-|\vec{k}|]}}{\sqrt{m^2+|\vec{k}|^2}}
\end{equation} 
is an asymptotic symbol of order $-1$ (see below and also \cite{RS} Chapter IX, example 7 and problem 68). Then because of Theorem \ref{wfsetasym},
\[WF(\Delta_{+}(x,m^2))\subseteq SP(\varphi)=\{ (0,\vec{0},-|\vec{k}|,\vec{k})\} \cup \{ (\pm |\vec{x}|,\vec{x},-\lambda|\vec{x}|,\mp \lambda \vec{x}) | \lambda >0 \} ,\]
 so the wavefront set of $\omega_2(x,0)$ must satisfy the same inclusion because it is just a continuous sum over distributions $\Delta_{+}(x,m^2)$ whose wavefront sets satisfy it (moreover the wave front sets do not depend on $m$). A sign convention comment is in order: we followed throughout the paper the definition of Fourier transform with $-$ sign, which is the one that is used to prove Theorem  \ref{wfsetasym} in the cited references. However, the $\mu$SC is stated assuming the opposite sign convention in the Fourier transform. This explains why the result above seems to be in contradicition with (\ref{microlocal}). Note also that the singular directions in momentum space are tangent to the light-cone, so the WF set is properly contained in the set used to define the $\mu$SC\footnote{An additional technical point should be clarified: we proved the microlocal spectrum condition property of $\omega_2(x,0)$, however in order to claim the same result for $\omega_2(x_1,x_2)$ we should consider the pullback of the wavefront set of $\omega_2(x,0)$ under $r:(x_1,x_2)\mapsto(x_1-x_2,0)$, with $x=x_1-x_2$ (see \cite[Theorem 8.2.3]{Hormander} for the behavior of the wavefront set under pullback). The pull-back under $r$ may enlarge the wavefront so that (\ref{microlocal}) does not hold, however since $r$ is a submersion we have $WF(\omega_2(x_1,x_2))=\left\{ (x_1,k;x_2,-k) \,\,|\,\, (x_1-x_2,k) \in WF(\omega_2(x,0)) \right\}$, see \cite{Strohmaier:2009zz} . }.    

In order to arrive to the above conclusion we claimed that $\phi$ is a phase function and $a$ an asymptotic symbol. The former is easy to justify, so we turn now to sketch the proof of the latter, which is not proven in \cite{RS} (it is left as an exercise). In order to see that $a(x,\vec{k},m)$ is an asymptotic symbol we present a possible approach, reducing to a 1-dimensional problem when we take derivatives: $\vec{k}\rightarrow k \in\mathbb{R}$. This is roughly justified because derivatives of $|\vec{k}|$ are of the form $k_i/|\vec{k}|$ and then behave like $|\vec{k}|^0$ at large $|\vec{k}|$.  First let us write  $a(x,\vec{k},m)=f(\vec{k})h(t,\vec{k})$, where $f(\vec{k})=(m^{2}+|\vec{k}|^{2})^{-\frac{1}{2}}$ and    $h(t,\vec{k})=\exp\{-itg(\vec{k})\}$ with $g(\vec{k})=(m^2+|\vec{k}|^2)^\frac{1}{2}-|\vec{k}|$. After some calculations one can probe that $f$ and $g$ are actually  asymptotic symbols of order $-1$, and for large $|\vec{k}|$,
 $|\partial_{k^{i}}^{\delta}h(t,\vec{k})|\leq |\vec{k}|^{-1-(\delta)}$ for all $\delta>0$. Applying Leibniz's rule we see that,

\begin{align*}
|\partial_{k}^{\beta} \partial_{t}^{\alpha}(f(k)h(t,k))| &\leq \sum_{0\leq \mu \leq \beta} {\beta \choose \mu} |\partial^{\mu}_{k}f(k)||\partial_{k}^{\beta-\mu }\partial_{t}^{\alpha}h(t,k)|\\
&=\sum_{0\leq \mu \leq \beta} {\beta \choose \mu} |\partial^{\mu}_{k}f(k)||\partial_{k}^{\beta-\mu }(g^{\alpha}(k)h(t,k))|\\
&=\sum_{0\leq \mu \leq \beta}\sum_{0\leq \nu \leq \beta-\mu} {\beta \choose \mu}{\beta-\mu \choose \nu} |\partial^{\mu}_{k}f(k)||\partial_{k}^{\nu }(g^{\alpha}(k)| |\partial_{k}^{\beta-\mu-\nu}h(t,k))|,
\end{align*}
Then if we separate the terms with $\nu=\beta-\mu$ from the others, we can bound them as $\leq C |k|^{-1-\alpha-\beta}$ for large $k$. The remaining therms, that is when $\beta-\mu-\nu$ is positive, can be bounded by $\leq C' |k|^{-2-\alpha-\beta}$,  using the comments above and the fact that, if $g$ is a symbol of order $-1$, then $g^{\alpha}$ is a symbol of order $-\alpha$ (see \cite{vestberg}, Lemma 3.6). Thus, we can drop them for large $k$. Finally,
\begin{equation*}
|\partial_{k}^{\beta} \partial_{t}^{\alpha}(f(k)h(t,k))|\leq C|k|^{-1-\alpha-\beta}\leq C|k|^{-1-\beta} \hspace{1cm} \text{  for large $k$}.
\end{equation*}
That is $a(t,k)$ is an asymptotic symbol of order $-1$.

\section*{Acknowledgments}

This work was supported by UBA and CONICET. We thank Gaston Giribet and  Shubho Roy for enlightening conversations. A. G. thanks the hospitality of ICTP while part of this work was done.

\end{document}